\documentclass[
,draft            
  ]
  {aipproc}

\layoutstyle{6x9}

\begin{document}

\title{Non-local Updates for Quantum Monte Carlo Simulations}

\author{Matthias Troyer}{
  address={Theoretische Physik, ETH Z\"urich, 8093 Z\"urich, Switzerland}
  ,altaddress={Computational Laboratory, ETH Z\"urich, 8092 Z\"urich, Switzerland} 
  ,email={troyer@phys.ethz.ch}
}

\author{Fabien Alet}{
  address={Theoretische Physik, ETH Z\"urich, 8093 Z\"urich, Switzerland}
  ,altaddress={Computational Laboratory, ETH Z\"urich, 8092 Z\"urich, Switzerland} 
  ,email={alet@phys.ethz.ch}
}

\author{Simon Trebst}{
  address={Theoretische Physik, ETH Z\"urich, 8093 Z\"urich, Switzerland}
  ,altaddress={Computational Laboratory, ETH Z\"urich, 8092 Z\"urich, Switzerland} 
  ,email={trebst@phys.ethz.ch}
}

\author{Stefan Wessel}{
  address={Theoretische Physik, ETH Z\"urich, 8093 Z\"urich, Switzerland}
  ,email={wessel@phys.ethz.ch}
}

\begin{abstract}
We review the development of update schemes for quantum lattice models simulated using world line quantum Monte Carlo algorithms. Starting from the Suzuki-Trotter mapping we discuss limitations of local update algorithms and highlight the main developments beyond Metropolis-style local updates: the development of cluster algorithms, their generalization to continuous time, the worm and directed-loop algorithms and finally a generalization of the flat histogram method of Wang and Landau to quantum systems.
\end{abstract}

\maketitle


\section{Quantum Monte Carlo World line algorithms }

Suzuki's realization in 1976  \cite{Suzuki} that the partition function of a $d$-dimensional quantum spin-$1/2$ system can be mapped onto that of a $(d+1)$-dimensional classical Ising model with special interactions enabled the straightforward simulation of arbitrary quantum lattice models, overcoming the restrictions of Handscomb's method \cite{Handscomb}. Quantum spins get mapped onto classical world lines and the Metropolis algorithm \cite{Metropolis} can be employed to perform local updates of the configurations. 

Just like classical algorithms the local update quantum Monte Carlo algorithm suffers from the problem of critical slowing down at second order phase transitions and the problem of tunneling out of metastable states at first order phase transitions. Here we review the development of non-local update algorithms, stepping beyond local update Metropolis schemes:
\begin{itemize}

\item 1993: the loop algorithm \cite{Evertz93}, a generalization of the classical cluster algorithms to quantum systems allows efficient simulations at second order phase transitions.

\item 1996: continuous time versions of the loop algorithm \cite{Beard96} and the local update algorithms \cite{Prokofev96} remove the need for an extrapolation in the discrete time step of the original algorithms (an approximation-free power-series scheme had been introduced for
the S=1/2 Heisenberg model already in \cite{Handscomb}, and a related, more general
method with local updates was presented in \cite{Sandvik91}).

\item from 1998: the worm algorithm \cite{Prokofev98}, the loop-operator \cite{Sandvik99,Dorneich} and the directed loop algorithms \cite{directedloop} remove the requirement of spin-inversion or particle-hole symmetry.

\item 2003: flat histogram methods for quantum systems \cite{Troyer03} allow efficient tunneling between metastable states at first order phase transitions.

\end{itemize}

\section{World lines and local update algorithms}

\subsection{The Suzuki-Trotter decomposition}
In classical simulations the Boltzmann weight of a configuration $c$ at an inverse temperature $\beta=1/k_BT$ is easily calculated from its energy $E_c$ as $\exp(-\beta E_c)$. Hence the thermal average of a quantity $A$
\begin{equation}
\langle A \rangle_{\rm classical} = \sum_c A_c \exp(-\beta E_c)/\sum_c \exp(-\beta E_c)
\end{equation}
can be directly estimated in a Monte Carlo simulation. The key problem for a quantum Monte Carlo simulation is that the simple exponentials of energies get replaced by  exponentials of the Hamilton operator $H$:
\begin{equation}
\langle A \rangle = {\rm Tr}\left[A\exp(-\beta H)\right]/{\rm Tr}\left[\exp(-\beta H)\right]
\end{equation}

The seminal idea of Suzuki \cite{Suzuki}, using a generalization of Trotter's formula \cite{Trotter}, was to split $H$ into two or more terms $H=\sum_i^N H_i$ so that the exponentials of each of the terms $\exp(-\beta H_i)$ is easy to calculate. Although the $H_i$ do not commute, the error in estimating the exponential
\begin{equation}
\exp(-\epsilon H)\approx\prod_i\exp(-\epsilon H_i)+{\cal O}(\epsilon^2)
\end{equation}
is small for small prefactors $\epsilon$ and better formulas of arbitrarily high order can be derived \cite{Suzukihigher}. Applying this approximation to the partition function we get Suzuki's famous mapping, here shown for the simplest case of two terms $H_1$ and $H_2$
\begin{eqnarray} 
Z&=&{\rm Tr}\left[\exp(-\beta H)\right]\;=\;
{\rm Tr}\left[\exp(-\Delta\tau(H_1+H_2)\right]^M\nonumber \\ 
\label{eq:suzuki} 
&=&{\rm Tr}\left[\exp(-\Delta\tau H_1)\exp(-\Delta\tau H_2)\right]^M+\;{\cal O} (\Delta \tau^2) \\
&=&\sum_{i_1,\ldots,i_{2M}}\langle i_{1}|U_{1}|i_{2}\rangle\langle
i_{2}|U_{2} |i_{3}\rangle\cdots \langle i_{2M-1}|U_{1}|i_{2M}\rangle
\langle i_{2M}|U_{2}|i_{1}\rangle\;+\;{\cal O} (\Delta\tau^2), \nonumber 
\end{eqnarray} 
where the time step is
$\Delta\tau\;=\;\beta/M$, the $|i_k\rangle$ each are complete orthonormal sets
of basis states, and the transfer matrices are $U_i=\exp(-\Delta\tau H_i)$.
The evaluation of the matrix elements
$\langle i|U_{1}|i'\rangle$ is straightforward since the $H_i$ are
chosen to be easily diagonalized.

\subsection{The World Line Representation}
\begin{figure}
\includegraphics{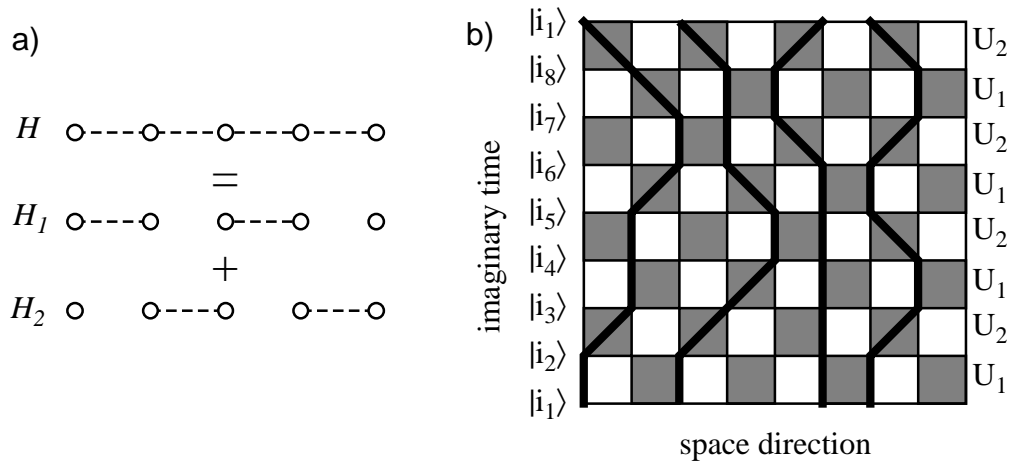}
\caption{
The ``checkerboard decomposition'': a) the Hamiltonian is split into odd and even bond terms. b) A graphical representation of Suzuki's mapping of a one-dimensional quantum system to a two-dimensional classical one, where an example world line configuration is shown.}
\label{fig:suzuki}
\end{figure}

As an example we consider a one-dimensional chain with nearest neighbor interactions. The Hamiltonian $H$ is split into odd and even bonds $H_1$ and $H_2$, as shown in Fig. \ref{fig:suzuki}a). Since the bond terms in each of these sums commute, the calculation of the exponential is easy. 
Equation (\ref{eq:suzuki}) can be interpreted as an evolution in
imaginary time (inverse temperature) of the
state $|i_1\rangle$ by the ``time evolution'' operators $U_{1}$ and $U_{2}$. 
Within each time interval $\Delta\tau$ 
the operators $U_1$ and and $U_2$ are each applied once. 
This leads to the famous ``checkerboard decomposition'', 
a graphical representation of the sum on a
square lattice, where the applications of the 
operators $U_i$ are marked by shaded squares
(see Fig.\ \ref{fig:suzuki}b). The configuration along each time slice 
corresponds to one of the states $|i_k\rangle$ in the sum 
(\ref{eq:suzuki}). 

This establishes the mapping of a one-dimensional quantum to a two-dimensional classical model where the four classical states at the corners of each plaquette interact with a four-site Ising-like interaction.  
For Hamiltonians with particle number (or magnetization) conservation we can take the mapping one step further. Since the conservation law applies locally on each shaded plaquette, particles on neighboring time slices can be connected and we  get a
representation of the configuration $\{|i_k\rangle\}$ in terms of world lines.
The sum over all configurations $\{|i_k\rangle\}$ with non-zero weights $\langle i_k|U|i_{k+1}\rangle$
corresponds to the sum over all possible world line configurations. In Fig. \ref{fig:suzuki}b) we show such a world line configuration for a model with one type of particle (e.g. a spin-1/2, hardcore boson or spinless fermion model). For models with more types of particles there
will be more kinds of world lines representing different particles (e.g. spin-up and spin-down fermions).

\subsection{Local Updates}

The world line representation can be used as a starting point of a quantum Monte Carlo algorithm \cite{Suzuki77}. Since particle number conservation prohibits the breaking of world lines, the local updates need to move world lines instead of just changing local states as in a classical model. 

\begin{figure}
\includegraphics[width=\textwidth]{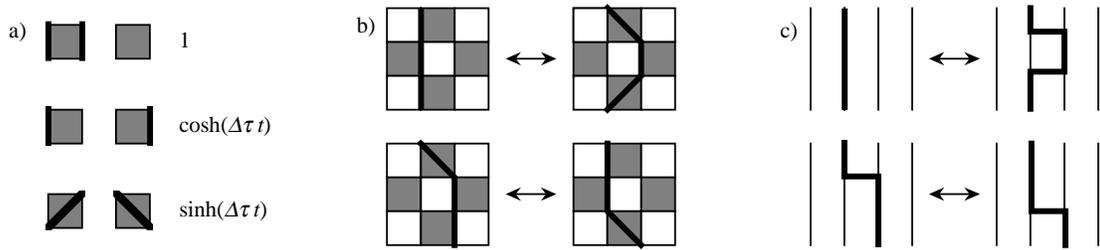}
\caption{
Examples of the two types of local moves used to update the world line
configuration in a tight-binding model with two states per site and Hamiltonian Eq. (\protect{\ref{eq:tight}}): a) plaquette weights $\langle i_k|U|i_{k+1}\rangle$ of the six possible local world line configurations in a tight binding model; b) the two types of updates in discrete time and c) in continuous time.}
\label{fig:localupdates}
\end{figure}

As an example we consider a one-dimensional tight binding model with Hamiltonian
\begin{equation}
H=-t\sum_i \left(c_i^\dag c_{i+1}+c_{i+1}^\dag c_{i}\right)\;,
\label{eq:tight}
\end{equation}
where $c_i^\dag$ creates a particle (spinless fermion or hardcore boson) at site $i$.  Fig. \ref{fig:localupdates}a shows the plaquette weights $\langle i_k|U|i_{k+1}\rangle$ for each of the six world line configurations on a shaded plaquette in this model. 

The local updates are quite simple and move a world line across a white plaquette  \cite{Suzuki77,Hirsch}, as shown in Fig. \ref{fig:localupdates}b). Slightly more complicated local moves are needed for higher-dimensional models \cite{Makivic92}, $t$-$J$ models \cite{Assaad, Troyer94} and Kondo lattice models \cite{Troyer94}.

Since these local updates cannot change global properties, such as the number of world lines or their spatial winding, they need to be complemented with global updates if the grandcanonical ensemble should be simulated \cite{Makivic92}. The problem of exponentially low acceptance rate of such moves was remedied only much later by the non-local update algorithms discussed below.
 
\subsection{The Continuous Time Limit}

The systematic error arising from the finite time step $\Delta\tau$ was originally controlled by an extrapolation to the continuous time limit $\Delta\tau\rightarrow0$ from simulations with different values of the time step $\Delta\tau$. It required a fresh look at quantum Monte Carlo algorithms by a Russian group \cite{Prokofev96} in 1996 to realize that, for a discrete quantum lattice model, this limit can already be taken during the construction of the algorithm and simulations can be performed directly at $\Delta\tau\rightarrow0$, corresponding to an infinite Trotter number $M=\infty$.

In this limit the Suzuki-Trotter formula  Eq. (\ref{eq:suzuki}) becomes equivalent to a time-dependent perturbation theory in imaginary time \cite{Prokofev96,Prokofev98}:
\begin{eqnarray}
\label{eq:pert}
Z&=&{\rm Tr}\exp(-\beta H)={\rm Tr}\left[\exp(-\beta H_0){\cal T}\exp\int_0^\beta d\tau V(\tau)\right], \\
&=&{\rm Tr}\left[\exp(-\beta H_0)\left(1-\int_0^\beta d\tau V(\tau)d\tau+\frac{1}{2}\int_0^\beta  d\tau_1\int_{\tau_1}^\beta  d\tau_2 V(\tau_1)V(\tau_2)+...\right)\right] \nonumber,
\end{eqnarray}
where the symbol ${\cal T}$ denotes time-ordering of the exponential.
The Hamiltonian $H=H_0+V$ is split into a diagonal term $H_0$ and an offdiagonal perturbation $V$. The time-dependent perturbation in the interaction representation is $V(\tau)=\exp(\tau H_0)V\exp(-\tau H_0)$. In the case of the tight-binding model the hopping term $t$ is part of the perturbation $V$, while additional diagonal potential or interaction terms would be  a part of $H_0$.

To implement a continuous time algorithm the first change in the algorithm is to keep only a list of times at which the configuration changes instead of storing the configuration at each of the $2M$ time slices in the limit $M\rightarrow\infty$. Since the probability for a jump of a world line [see Fig. \ref{fig:localupdates}a)] and hence a change of the local configuration is $\sinh(\Delta\tau t) \propto \Delta\tau\propto 1/M$ the number of such changes remains finite in the limit $M\rightarrow\infty$. The representation is thus well defined,
and, equivalently, in Eq. (\ref{eq:pert}) only a finite number of terms contributes in a finite system.

The second change concerns the updates, since the probability for the insertion of a pair of jumps in the world line [the upper move in Fig. \ref{fig:localupdates}b)] vanishes as
\begin{equation}
P_{\mbox{insert jump}} = \sinh^2(\Delta\tau t)/\cosh^2(\Delta\tau t) \propto \Delta\tau^2\propto1/M^2\rightarrow 0
\end{equation}
in the continuous time limit. To counter this vanishing probability, one proposes to insert a pair of jumps not at a specific location but {\it anywhere} inside a finite time interval \cite{Prokofev96}. The integrated probability then remains finite in the limit $\Delta\tau\rightarrow 0$. Similarly instead of shifting a jump by $\Delta\tau$ [the lower move in Figs. \ref{fig:localupdates}b,c)] we move it by a finite time interval in the continuous time algorithm.

\subsection{Stochastic Series Expansion}

An alternative Monte Carlo algorithm, which also does not suffer from time discretization, is the stochastic series expansion (SSE) algorithm \cite{Sandvik91}, a generalization of Handscomb's algorithm \cite{Handscomb} for the Heisenberg model. It starts from a Taylor expansion of the partition function in orders of $\beta$:
\begin{eqnarray}
\label{eq:sse}
Z&=&{\rm Tr}\exp(-\beta H)=\sum_{n=0}^\infty\frac{\beta^n}{n!}{\rm Tr}(-H)^n \nonumber \\
&=&\sum_{n=0}^\infty\frac{\beta^n}{n!}\sum_{\{i_1,...i_n\}}\sum_{\{b_1,...b_n\}}\langle i_1|-H_{b_1}|i_2\rangle\langle i_2|-H_{b_2}|i_3\rangle\cdots\langle i_n|-H_{b_n}|i_1\rangle 
\end{eqnarray}
where in the second line we decomposed the Hamiltonian $H$ into a sum of single-bond terms $H=\sum_b H_b$, and again inserted complete sets of basis states. We end up with a similar representation as Eq. (\ref{eq:suzuki}) and a related world-line picture with very similar update schemes. For more details of the SSE method we refer to the contribution of A.W. Sandvik in this proceedings volume.

\begin{figure}
\includegraphics[width=\textwidth]{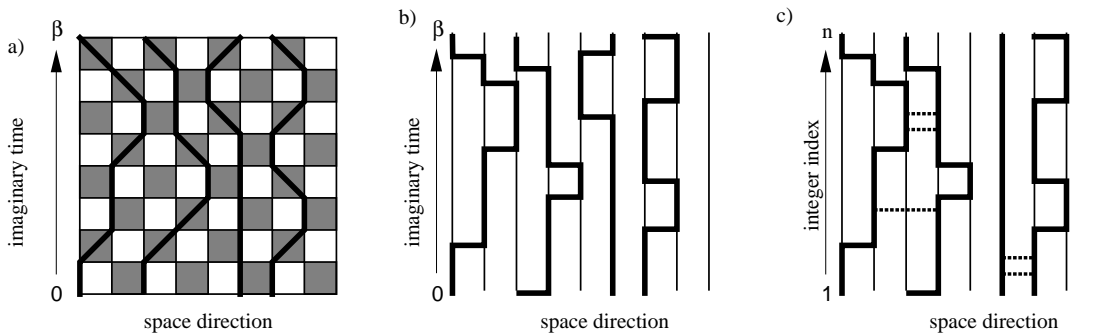}
\caption{
A comparison of a) world lines in discrete time, b) in continuous time and c) a similar configuration in the SSE representation. In the SSE representation the continuous time index is replaced by an integer order index of the operators, at the cost of additional diagonal terms (the dashed lines).}
\label{fig:sse}
\end{figure}

The SSE representation can be formally related to the world line representation by observing that  Eq. (\ref{eq:sse}) is obtained from Eq. (\ref{eq:pert}) by setting $H_0=0$, $V=H$ and integrating over all times (compare also Fig. \ref{fig:sse}) $\tau_i$ \cite{Sandvik97}. This mapping also shows the advantages and disadvantages of the two representations. The SSE representation corresponds to a perturbation expansion in all terms of the Hamiltonian, whereas world line algorithms treat the diagonal terms in $H_0$ exactly and perturb only in the offdiagonal terms $V$ of the Hamiltonian. World line algorithms hence need only fewer terms in the expansion, but pay for it by having to deal with imaginary times $\tau_i$. The SSE representation is thus preferred except for models with large diagonal terms (e.g. bosonic Hubbard models) or for models with time-dependent actions (e.g. dissipative quantum systems \cite{CaldeiraLegget}). 

\section{The loop algorithm}
While the local update world line and SSE algorithms enable the simulation of quantum systems  they suffer from critical slowing down at second order phase transitions. Even worse, changing the spatial and temporal winding numbers has an exponentially small acceptance rate. While the restriction to zero spatial winding can be viewed as a boundary effect, changing the temporal winding number and thus the magnetization or particle number is essential for simulations in the grand canonical ensemble. 

The solution to these problems came with the loop algorithm \cite{Evertz93} and its continuous time version \cite{Beard96}. These algorithms, generalizations of the classical cluster algorithms \cite{SwendsenWang} to quantum systems, not only solve the problem of critical slowing down, but also updates the winding numbers efficiently for those systems to which it can be applied.

 \begin{figure}[b]
\includegraphics[width=10cm]{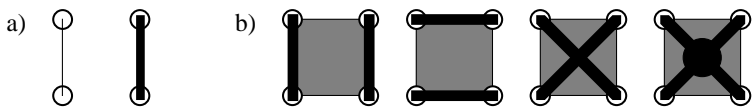}
\caption{
a) in the cluster algorithms for classical spins two sites can either be connected (thick line) or disconnected (thin line). b) in the loop algorithm for quantum spins two or fours spins on a shaded plaquette must be connected.}
 \label{fig:clusters}
\end{figure}

Since there is an extensive recent review of the loop algorithm \cite{Evertz03}, we will only mention the main idea behind the loop algorithm here. In the classical Swendsen-Wang cluster algorithm each bond in the lattice is considered, and with a probability depending on the local configuration two neighboring spins are either ``connected'' or left ``disconnected'', as shown in Fig. \ref{fig:clusters}a).  ``Connected'' spins form a cluster and must be flipped together. Since the average extent of these cluster is just the correlation length of the system, updates are performed on physically relevant length scales and autocorrelation times are substantially reduced.

 \begin{figure}
\includegraphics[width=\textwidth]{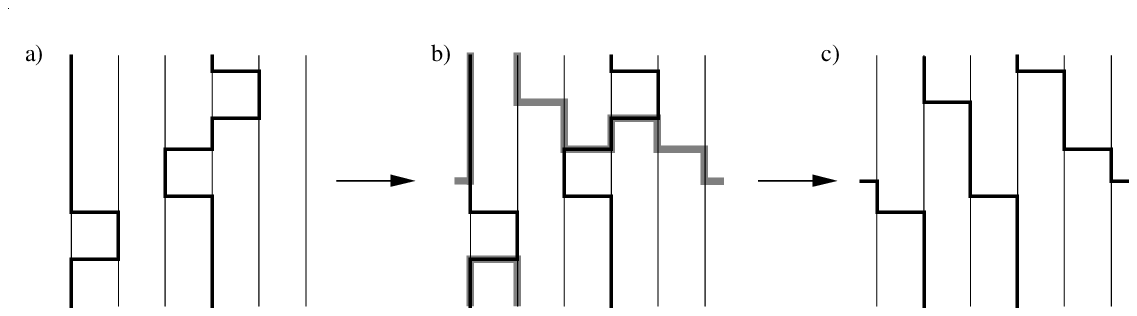}
\caption{
A loop cluster update: a) world line configuration before the update, where the world line of a particle (or up-spin in a magnetic model) is drawn as a thick line and that of a hole (down-spin) as a thin line; b) world line configuration and a loop cluster (grey line); c) the world line configurations after all spins along the loop have been flipped.
}
\label{fig:loopupdate}
\end{figure}

Upon applying the same idea to world lines in QMC we have to take into account that (in systems with particle number or magnetization conservation) the world lines may not be broken. This implies that a single spin on a plaquette cannot be flipped by itself, but at least two, or all four spins must be flipped in order to create valid updates of the world line configurations. Instead of the two possibilities ``connected'' or ``disconnected'', four connections are possible on a plaquette, as shown in Fig. \ref{fig:clusters}b): either horizontal neighbors, vertical neighbors, diagonal neighbors or all four spins might be flipped together. The specific choices and probabilities depend, like in the classical algorithm, on details of the model and the world line configuration. Since each spin is connected to two (or four) other spins, the cluster has a loop-like shape (or a set of connected loops), which is the origin of the name ``loop algorithm'' and is illustrated in Fig. \ref{fig:loopupdate}. 

While the loop algorithm was originally developed only for six-vertex and spin-1/2 models \cite{Evertz93} it has been generalized to higher spin models \cite{higherspin}, anisotropic spin models \cite{anisotropicspin}, Hubbard \cite{hubbard} and $t$-$J$ models \cite{Ammon}.

\subsection{Applications of the loop algorithm}

Out of the large number of applications of the loop algorithm we want to mention only a few which highlight the advances made possible by the development of this algorithm and refer to Ref. \cite{Evertz03} for a more complete overview.
\begin{itemize}
\item The first application of the discrete and continuous time loop algorithms \cite{Ying92,Beard96} were high accuracy simulations of the ground state parameteres of the square lattice Heisenberg antiferromagnet, establishing beyond any doubt the existence of N\'eel order even for spin $S=1/2$.

\item The exponential divergence of the correlation length in the same system could be studied on much larger systems with up to one million spins \cite{Kim97,Kim98,Beard98} and with much higher accuracy than in previous simulations \cite{Makivic92}, investigating not only the leading exponential behavior but also higher order corrections.

\item For quantum phase transitions in two-dimensional quantum Heisenberg antiferromagnets, simulations using local updates had been restricted to small systems with up to 200 spins at not too low temperatures and had given contradicting results regarding the universality class of the phase transitions \cite{Sandvik94,Katoh94}. The loop algorithm enabled simulations on up to one hundred times larger systems at ten times lower temperatures, allowing the accurate determination of the critical behavior at quantum phase transitions \cite{Troyer97,Sandvik00}. 

\item Similarly, in the two-dimensional quantum $XY$ model the loop algorithm allowed accurate simulations of the Kosterlitz-Thouless phase transition \cite{Harada97}, again improving on results obtained using local updates \cite{Makivic92b}.

\item In SU(4) square lattice antiferromagnets, the loop algorithm could clarify that a spin liquid state thought to be present based on data obtained using local update algorithms on small lattices \cite{Santoro} is actually N\'eel ordered \cite{Harada03}.

\item A generalization, which allows to study infinite systems in the absence of long range order, was invented \cite{Evertz01}.

\item The meron cluster algorithm, an algorithm based on the loop algorithm, solves the negative sign problem in some special systems \cite{Wiese99}.

\end{itemize}

\section{Worm and directed loop algorithms}
\subsection{Problems of the loop algorithm in a magnetic field}
As successful as the loop algorithm is, it is restricted -- as the classical cluster algorithms -- to models with spin inversion symmetry (or particle-hole symmetry). Terms in the Hamiltonian which break this spin-inversion symmetry -- such as a magnetic field in a spin model or a chemical potential in a particle model -- are not taken into account during loop construction. Instead  they enter through the acceptance rate of the loop flip, which can be exponentially small at low temperatures.

 \begin{figure}
\includegraphics[height=5cm]{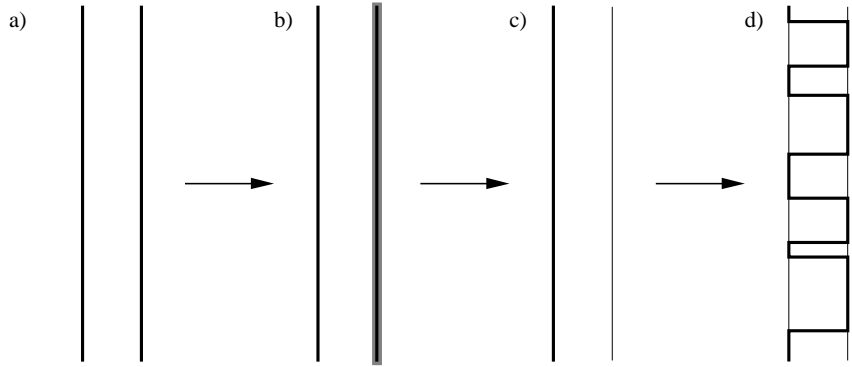}
\caption{
A loop update for two antiferromagnetically coupled spins in a magnetic field with $J=h$. a) Starting from the triplet configuration $|\uparrow\uparrow\rangle$, b) a loop is constructed, proposing to go to c), the intermediate configuration $|\uparrow\downarrow\rangle$, which has an exponentially small acceptance rate, and finally into configurations like d) which represent the singlet state $1/\sqrt{2}(|\uparrow\downarrow\rangle-|\downarrow\uparrow\rangle)$. As in the previous figure a thick line denotes an up-spin and a thin line a down-spin.
}
\label{fig:loopfield}
\end{figure}

 As an example consider two $S=1/2$ quantum spins in a magnetic field:
\begin{equation}
H=J{\bf S}_1{\bf S}_2-h(S_1^z+S_2^z)
\end{equation}
In a field $h=J$ the singlet state $1/\sqrt{2}(|\uparrow\downarrow\rangle-|\downarrow\uparrow\rangle)$ with energy $-3/4J$ is degenerate with the triplet state $|\uparrow\uparrow\rangle$ with energy $1/4J-h=-3/4J$, but he loop algorithm is exponentially inefficient at low temperatures. As illustrated in Fig. \ref{fig:loopfield}a), we start from the triplet state 
 $|\uparrow\uparrow\rangle$  and propose a loop shown in Fig. \ref{fig:loopfield}b). The loop construction rules, which do not take into account the magnetic field, propose to flip one of the spins and go to the intermediate configuration $|\uparrow\downarrow\rangle$ with energy $-1/4J$ shown in Fig. \ref{fig:loopfield}c). This move costs potential energy $J/2$ and thus has an {\it exponentially small acceptance rate} $\exp(-\beta J/2)$. Once we accept this move, immediately many small loops are built, exchanging the spins on the two sites, and gaining exchange energy $J/2$ by going to the spin singlet state. A typical world line configuration for the singlet is shown in Fig. \ref{fig:loopfield}d). The reverse move has the same exponentially small probability, since the probability to reach a world line configuration without any exchange term [Fig. \ref{fig:loopfield}c)] from a spin singlet configuration [Fig. \ref{fig:loopfield}d)] is exponentially small.

This example clearly illustrates the reason for the exponential slowdown: in a first step we {\it lose all potential energy}, before {\it  gaining it back in exchange energy}. A faster algorithm could thus be built if, instead of doing the trade in one big step, we could trade potential with exchange energy in small pieces, which is exactly what the worm algorithm does.

\subsection{The Worm Algorithm}

\begin{figure}
\includegraphics[height=5cm]{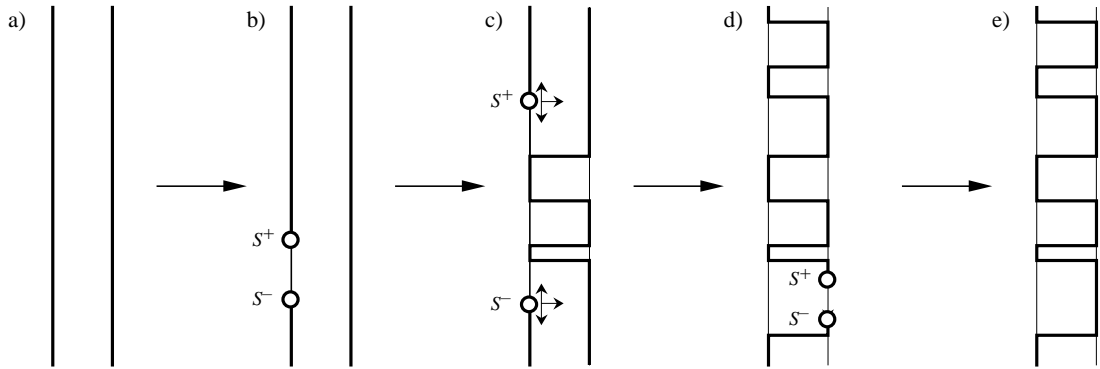}
\caption{
A worm update for two antiferromagnetically coupled spins in a magnetic field with $J=h$. a) starting from the triplet configuration $|\uparrow\uparrow\rangle$ a worm is constructed in b) by inserting a pair of $S^+$ and $S^-$ operators. c) these ``worm end'' operators are then moved by local updates until d) they meet again, when a move to remove them is proposed, which leads to the closed world line configuration e). As in the two previous figures a thick line denotes an up-spin and a thin line a down-spin.
}
\label{fig:worm}
\end{figure}

The worm algorithm \cite{Prokofev98} works in an extended configuration space, where in addition to closed world line configurations one open world line fragment (the ``worm'') is allowed. Formally this is done by adding a source term to the Hamiltonian which for a spin model is
\begin{equation}
H_{\rm worm}=H-\eta\sum_i(S_i^++S_i^-)\;.
\end{equation}
This source term allows world lines to be broken with a matrix element proportional to $\eta$. The worm algorithm now proceeds as follows: a worm (i.e. a world line fragment) is created by inserting a pair $( S_i^+,S_i^-)$ of operators at nearby times, as shown in Fig. \ref{fig:worm}a,b). The ends of this worm are then moved randomly in space and time [Fig. \ref{fig:worm}c)], using local Metropolis or heat bath updates until the two ends of the worm meet again as in Fig. \ref{fig:worm}d). Then an update which removes the worm is proposed, and if accepted we are back in a configuration with closed world lines only, as shown in Fig. \ref{fig:worm}e).
This algorithm is straightforward, consisting just of local updates of the worm ends in the extended configuration space but it can perform nonlocal changes. A worm end can wind around the lattice in the temporal or spatial direction and that way change the magnetization and winding number. 

In contrast to the loop algorithm in a magnetic field, where the trade between potential and exchange energy is done by first losing all of the potential energy, before gaining back the exchange energy, the worm algorithm performs this trade in small pieces, never suffering from an exponentially small acceptance probability.
While not being as efficient as the loop algorithm in zero magnetic field (the worm movement follows a random walk while the loop algorithm can be interpreted as a self-avoiding random walk), the big advantage of the worm algorithm is that it remains efficient in the presence of a magnetic field.

A similar algorithm was already proposed more than a decade earlier \cite{Cullen83}. Instead of a random walk using fulfilling detailed balance at every move of the worm head in this earlier algorithm just performed a random walk. The {\it a posteriori} acceptance rates are then often very small and the algorithm is not efficient, just as the small acceptance rates for loop updates in magnetic fields make the loop algorithm inefficient. This highlights the importance of having the cluster-building rules of a non-local update algorithm closely tied to the physics of the problem.

\subsection{The Directed Loop Algorithm}

Algorithms with a similar basic idea are the operator-loop update \cite{Sandvik99,Dorneich} in the SSE formulation and the directed-loop algorithms \cite{directedloop} which can be formulated in both an SSE and a world-line representation. Like the worm algorithm, these algorithms create two world line discontinuities, and move them around by local updates. The main difference to the worm algorithm is that here these movements do not follow an unbiased random walk but have a preferred direction, always trying to move away from the last change. 
The directed loop algorithms might thus be more efficient than the worm algorithm but no direct comparison has been performed so far. 
For more details see the contribution of A.W. Sandvik in this volume.

\subsection{Applications}

Just as the loop algorithm enabled a break-through in the simulation of quantum magnets in zero magnetic field, the worm and directed loop algorithms allowed simulations of bosonic systems with better efficiency and accuracy. A few examples include:

\begin{itemize}
\item Simulations of quantum phase transitions in soft-core bosonic systems, both for uniform models \cite{Prokofev98} and in magnetic traps \cite{Prokofev02}.
\item By being able to simulate substantially larger latttices than by local updates \cite{Batrouni} the existence of supersolids in hard-core boson models was clarified \cite{Hebert01} and the ground-state \cite{Hebert01,Bernardet02} and finite-temperature phase diagrams \cite{Schmid} of two-dimensional hard-core boson models have been determined.
\item Magnetization curves of quantum magnets have been calculated \cite{Kashurnikov99}.
\end{itemize}

\section{Flat Histograms and First Order Phase Transitions}

The main problem during the simulation of a first order phase transition is the exponentially slow tunneling time between the two coexisting phases. For classical simulations the multi-canonical algorithm \cite{Berg92} and recently the Wang-Landau algorithm \cite{WangLandau} eases this tunneling by reweighting configurations such as to achieve a ``flat histogram'' in energy space. In a canonical simulation the probability of visiting an energy level $E$ is $\rho(E)p(E)\propto\rho(E)\exp(-\beta E)$ where the density of states $\rho(E)$ is the number of states with energy $E$. While the multi-canonical algorithm \cite{Berg92} changes the canonical distribution $p(E)$ by reweighting it in an energy-dependent way, the algorithm by Wang and Landau discards the notion of temperature and directly uses the density of states to set $p(E)\propto1/\rho(E)$, which gives a constant probability in energy space $\rho(E)p(E)={\rm const.}$. The unknown quantity $\rho(E)$ is determined self-consistently in an iterative way and then allows to directly calculate the free energy
\begin{equation}
F = -k_B T \ln\sum_E\rho(E)\exp(-\beta E)
\end{equation}
and other thermodynamic quantities at any temperature. The main change to a simulation program using a canonical distribution is to replace the canonical probability $p(E)=\exp(-\beta E)$ by the inverse density of states $p(E)=1/\rho(E)$.

This algorithm cannot be straightforwardly used for quantum systems, since the density of states $\rho(E)$ is not directly accessible for those. Instead we recently proposed \cite{Troyer03} to start from the SSE formulation of the partition function Eq. (\ref{eq:sse}):
\begin{eqnarray}
\label{eq:qwl}
F &=& -k_BT \ln {\rm Tr}\exp(-\beta H)=-k_BT \ln \sum_{n=0}^\infty\frac{\beta^n}{n!}{\rm Tr}(-H)^n \nonumber \\  &=& -k_BT \ln \sum_{n=0}^\infty\frac{\beta^n}{n!}\sum_{\{i_1,...i_n\}}\sum_{\{b_1,...b_n\}}\langle i_1|-H_{b_1}|i_2\rangle\langle i_2|-H_{b_2}|i_3\rangle\cdots\langle i_n|-H_{b_n}|i_1\rangle \nonumber \\
&\equiv& -k_BT \ln \sum_{n=0}^\infty\beta^n g(n).
\end{eqnarray}
The coefficient $g(n)$ of the $n$-th order term in an expansion in the inverse temperature $\beta$ now plays the role of the density of states $\rho(E)$ in the classical algorithm. Similar to the classical algorithm, by  using $1/g(n)$ as the probability of a configuration instead of the usual SSE weight, a flat histogram in the order $n$ of the series is achieved. Alternatively instead of such a high-temperature expansion a finite-temperature perturbation series can be formulated \cite{Troyer03}.

This algorithm was shown to be effective at first order phase transitions in quantum systems and promises to be effective also for the simulation of quantum spin glasses.

\section{Which algorithm is the best?}
Since there is no ``best algorithm'' suitable for all problems we conclude with a guide on how to pick the best algorithm for a particular problem.
\begin{itemize}
\item For models with particle-hole or spin-inversion symmetry a loop algorithm is optimal \cite{Evertz93,Beard96,Sandvik99}. Usually an SSE representation \cite{Sandvik99} will be preferred unless the action is time-dependent (such as long-range in time interactions in a dissipative quantum system) or there are large diagonal terms, in which case a world line representation is better.
\item For models without particle hole symmetry a worm or directed-loop algorithm is the best choice:
\begin{itemize}
\item if the Hamiltonian is diagonally dominant use a worm \cite{Prokofev98} or directed loop \cite{directedloop} algorithm in a world line representation.
\item otherwise ause  directed-loop algorithm in an SSE representation. \cite{Sandvik99,Dorneich,directedloop}.
\end{itemize}

\item At first order phase transition a generalization of Wang-Landau sampling to quantum systems  should be used \cite{Troyer03}.
\end{itemize}

The source code for some of these algorithms is available on the Internet. Sandvik has published a FORTRAN version of an SSE algorithm for quantum magnets \cite{SandvikCode}. The ALPS ({\bf A}lgorithms and {\bf L}ibaries for {\bf P}hysics {\bf S}imulations) project is an open-source effort to provide libraries and application frameworks for classical and quantum lattice models as well as C++ implementations of the loop, worm and directed-loop algorithms \cite{ALPS}.


\begin{theacknowledgments}
We acknowledge useful discussions with H.G. Evertz, N. Kawashima, N. Prokof'ev and  A.~Sandvik about the relationship between the various cluster-update algorithms for quantum systems. F.A, S.T and S.W acknowledge support of the Swiss National Science Foundation.
\end{theacknowledgments}



\end{document}